\documentclass[twocolumn,showpacs,preprintnumbers,amsmath]{revtex4}

\usepackage[pdftex]{graphicx}
\usepackage{dcolumn}
\usepackage{bm}
\usepackage[english]{babel}

\newcommand{\ddinero}{\Delta w}
\newcommand{\round}[1]{\text{\emph{round}}(#1)}
\newcommand{\promedio}[1]{\left < #1 \right >}

\newcommand{\pentera}[1]{\left \lfloor #1 \right \rfloor}

\begin{document}

 \title{Class formation in a social network with asset exchange.}
 \author{Christian H. Sanabria Monta\~na}
 \email{sanabria@mda.cinvestav.mx}
 \author{Rodrigo Huerta-Quintanilla.}
 \email{rhuerta@mda.cinvestav.mx}
 \affiliation{
Centro de Investigaciones y de Estudios Avanzados del Instituto Polit\'ecnico Nacional\\ 
Unidad M\'erida, Departamento de F\'isica Aplicada.\\
Km 6 carretera antigua a progreso, M\'erida, Yucat\'an.
}
\author{Manuel Rodr\'iguez-Achach.}
 \email{achachm@gmail.com}
 \affiliation{
Universidad Veracruzana, Departamento de F\'isica, Facultad de F\'isica e Inteligencia Artificial\\
Circuito G. Aguirre Beltran s/n, Zona Universitaria Xalapa, Veracruz.
}

 \date{\today}

\begin{abstract}
 \noindent We study two kinds of economic exchange, additive and multiplicative, 
in a system of $N$ agents. The work is divided in two parts, in the first one, 
the agents are free to interact with each other. The system evolves to a Boltzmann-Gibbs 
distribution with additive exchange and condenses with a multiplicative one. If bankruptcy
is introduced, both types of exchange lead to condensation. Condensation times have been studied. 
In the second part, the agents are placed in a social network. We analyze the behavior of wealth distributions in time, and the formation of economic classes was observed for certain values of
network connectivity.
\end{abstract}


\maketitle

\section{\label{sec:level1}Introduction}
\begin{figure}[!htp]
 \centering
 \includegraphics[scale=0.35, angle=0]{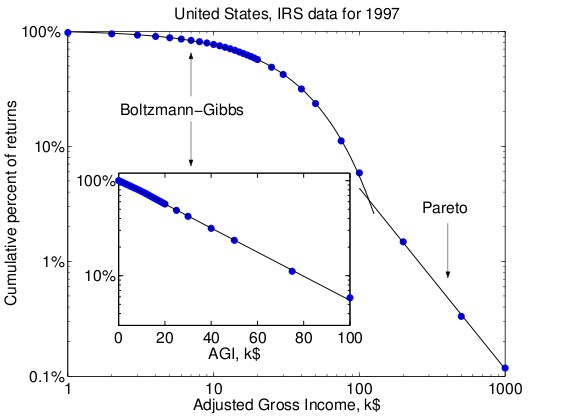}
  \caption{\label{fig:intro} Cumulative probability distribution of US individual income for 1997 (Figure (6) from \cite{art:15}).}
\end{figure}

\noindent We have had in the past years a large amount of literature dealing 
with the study of the distribution of wealth in agent based models with 
various kinds of interaction rules \cite{art:15,art:5,art:8}. 
Several distributions such as \emph{Boltzmann - Gibbs}, \emph{Gamma}, or \emph{Pareto} 
can be obtained according to the different conditions of the models 
\cite{art:3,art:6,art:10}. It is well known that real data analysis from
several countries \cite{book:10} yield a \emph{Boltzmann-Gibbs} distribution for that sector of the
population with lowest wealth, who are the majority, and a \emph{Pareto} 
distribution for the minority of the population with the highest values of wealth (see Figure \ref{fig:intro} for an example).

\noindent This particular behavior has been reproduced, to some extent, using different 
kinds of assumptions \cite{art:13,art:17}.

\noindent In this article we describe a model where this behavior
is obtained and the appearance of social classes is observed.
The model is agent--based, with agents arranged into a social network, 
 and we define very simple rules for wealth exchange. We analyze 
the conditions under which the different types of distributions are 
obtained and also the conditions for the social classes to appear.

\noindent In the next section of the paper we describe the interacting model for 
several cases under study. We focus the analysis on the behavior of entropy, 
poverty (defined as the minimum of wealth, where the 
amount of money  is less than the minimum allowed exchange) and wealth 
distributions in sections III and IV. 

\noindent In section III the dynamics of the model with and without
bankruptcy (defined as the impossibility to participate in a trade due to the wealth is less than the minimum allowed exchange) is investigated. The agents are not placed in a network, which is 
equivalent to a system where the agents are placed in a fully connected network. 
In section IV we study the effect of a non-fully connected network. Finally in the last 
section we give our conclusions.

\section{Interacting model}\label{sec:2}
\subsection{Wealth exchange models}\label{sub:2:1}
\noindent The kind of wealth exchange models that we have considered in the present 
work are those in which the total amount of wealth before and after the 
interaction is conserved, also known as elastic collision models. 
That is, if ($i$,$j$) are the labels of the two agents involved, and 
their wealths are ($w_i$,$w_j$) respectively, then we can write
\begin{align}
 w_i(t + \Delta t) = w_i(t) + \ddinero \label{eq:1},\\
 w_j(t + \Delta t) = w_j(t) - \ddinero \label{eq:6}.
\end{align}
Since wealth is preserved in the interaction, we have that 
$w_i(t+\Delta t) + w_j(t+\Delta t) = w_i(t) + w_j(t)$. It is important to remark that agents are not allowed 
to have negative wealth. Also is necessary to mention that agent $i$ has a probability $p$ to lose $\ddinero$, while 
agent $j$ has a probability $1-p$. In our model the agents have the same probability to win or to lose, then the value of $p$ is $0.5$. Furthermore the agents have only integer values of wealth, hence, $\ddinero$ must be an integer value. 
 
\noindent According to which amount we choose for $\ddinero$, we can 
have two distinct processes: additive or multiplicative exchange. In the additive exchange we have 
$$\ddinero = const\text{,}$$ 
this means that the exchange money is fixed in time and is independent of wealth \cite{art:1}. In the multiplicative case, also known as \emph{Yard-Sale} model \cite{art:11}, 
we have taken $\ddinero$ as 
$$\ddinero = \round{\nu\cdot\min(w_i(t),w_j(t))}\text{,}$$ 
where we used $\round{x}$ as the integer closest to $x$, and $0 < \nu < 1$. One difference between the multiplicative case and the additive case is the dependence on time for $\ddinero$, which in the latter case it is not shown. 

\subsection{Interacting model with no network, or undirected fully connected 
network}\label{sub:2:2}
\noindent We have considered a closed population composed of $N$ agents, where 
each of them has the possibility to exchange her wealth through the exchange
rules mentioned above, with any other agent in the population.

\noindent Based on the aforementioned we can establish the equivalence between 
a population with no network and a population embedded in an undirected fully 
connected network, where each agent $i$ has a connectivity $Z^i$ \cite{book:12} 
equal to $N-1$. Therefore equations (\ref{eq:1}) and (\ref{eq:6}) are valid 
for all ($i$,$j$) that belongs to the population.

\subsection{Interacting model with a undirected non-fully connected network}\label{sub:2:3}
\noindent We introduce an undirected  network in the population, in which every agent $i$ has 
$k_i$ links to other agents, such that $1 \leq k_i \leq k_{max}$, 
where $k_{max}$ is the highest possible number of links allowed in the dynamics.
In other words we have used a random network with an arbitrary degree distribution such that $p(k)$ is given by:

\begin{equation}\label{eq:7}
p(k) = \left \{ \begin{array}{lr}
                \frac{1}{k_{max}}; & k \in [1,k_{max}]\\
                \\
                0; &  \text{otherwise}
               \end{array}\right.
\end{equation}
\noindent We used this distribution because we want to have control over the 
maximum number of links for each agent, and therefore see the effect that
this parameter has on the wealth distribution.

\noindent The average number of links $\promedio{k}$ (or the connectivity 
mean value $\promedio{Z}$) in the network is given by
\begin{equation}\label{eq:2}
 \promedio{Z} = \promedio{k} = \frac{1}{N}\sum_{i=1}^{N} k_i = \sum_{k=1}^{k_{max}} p(k)\cdot k = \frac{k_{max}+1}{2}.
\end{equation}

\noindent With this in mind the dynamics in the network is of course 
different from case \ref{sub:2:2}, since agent $i$ will only interact with 
her $k_i$ links through equations (\ref{eq:1}) and (\ref{eq:6}).\\

\noindent One big difference between the fully connected network and the non-fully connected network is the connectivity. In a fully connected network the connectivity $Z^i$ is the same for all agents and is equal to $N-1$, meanwhile in a non-fully connected network the connectivity $Z^i$ is in the $[1,k_{max}]$ interval.

\subsection{Computer simulations}\label{sub:2:4}
\noindent The simulation begins with a population of $N$ agents. Every agent 
has the same initial amount of wealth $\promedio{w}$. To simulate an undirected fully connected network every agent has been linked with the other $N-1$ agents having in total $N(N-1)/2$ links on the network. To simulate an undirected non-fully connected network, we build it in the following way:
\begin{enumerate}
 \item A number of links $k$ is given randomly to every agent, using a uniform distribution at the interval $[1,k_{max}]$. In such a way that we reproduce $p(k)$ as in equation (\ref{eq:7}).
 \item An agent $i$ chooses randomly other agent $j$, and they will stablish a link as long as the agent $j$ has not completed its $k_j$ links assigned; otherwise the agent $i$ will choose other agent to connect. This is repeated until agent $i$ has completed his $k_i$ links.
 \item The step $2$ is repeated for every agent in the network.
\end{enumerate} 
Once the network has been built, the next part is to consider temporal evolution of system by Monte Carlo steps (MCS).
 
\noindent In a Monte Carlo step (MCS or time step), two agents ($i$,$j$) are chosen. 
In the case of undirected fully connected network, both agents 
are chosen randomly from the whole population; whereas in an undirected 
non-fully connected network, the first agent $i$ is chosen randomly, but the second agent $j$ is only chosen randomly from the links that agent $i$ has. 

\noindent The agents interact through a fair bet, that is, the probability 
to win is $p=0.5$. An amount of wealth $\ddinero$ is transferred from the 
loser to the winner. If the wealth of the loser is less than 
$\ddinero$, the transaction is not done.

\noindent In all simulations we have used $N=500$, $\promedio{w} = 100$ and 
$10^2$ realizations during $4\times10^5$ MCS. In the additive exchange the 
value of $c$ is $20$, in the multiplicative exchange $\nu = 0.2$. Within a 
network, the average number of links (connectivity mean value) obtained by 
the computer simulations were $1.468$, $1.952$, $2.428$ and $10.06$ 
(these values correspond to values of $k_{max}=2,3,4$ and $20$, respectively).

\section{Dynamics with undirected fully connected 
network}
\subsection{Exchange without bankruptcy}
\begin{figure}[!htp]
 \centering
 \includegraphics[scale=0.3, angle=0]{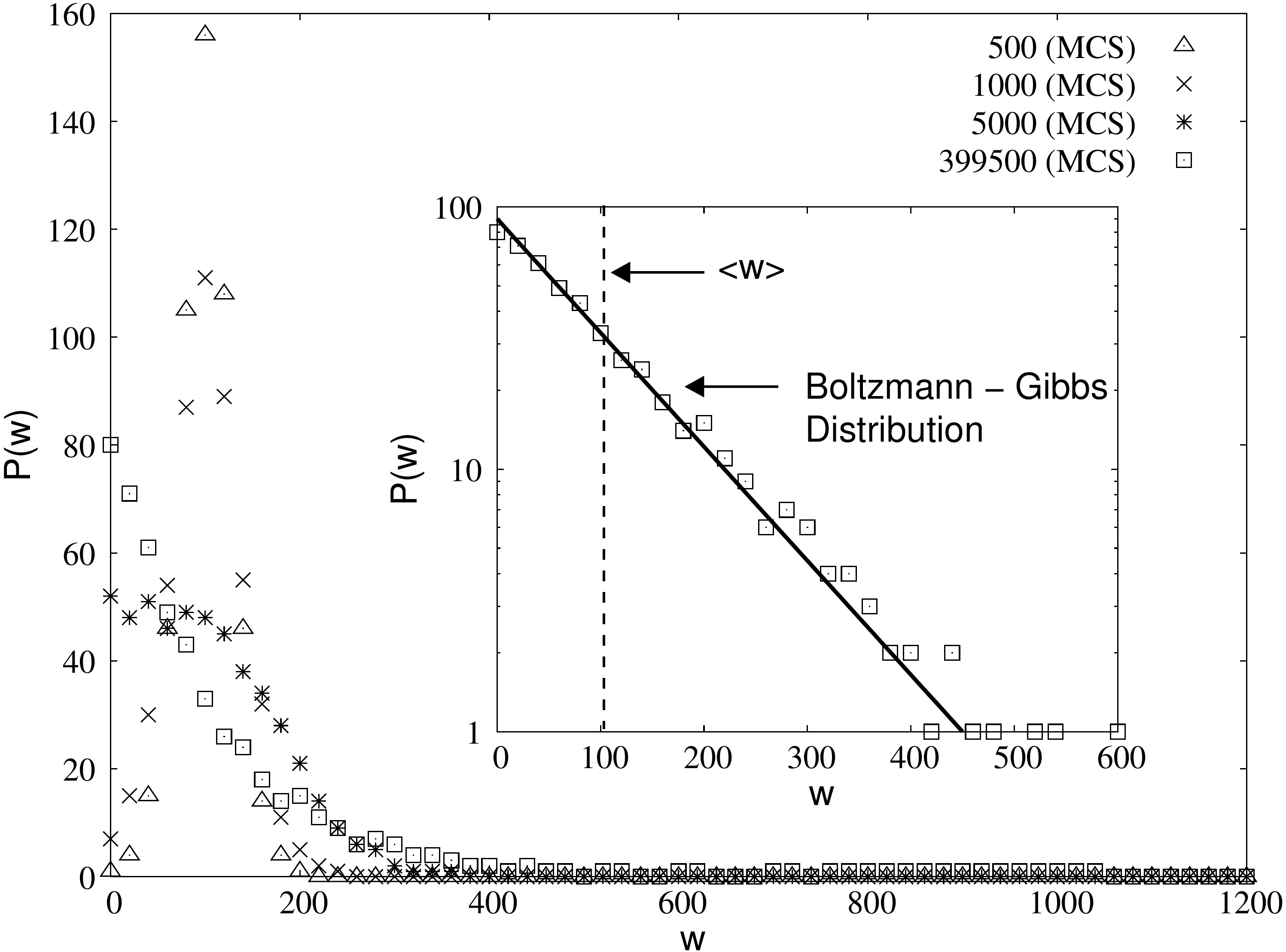}
 \caption{\label{fig:1a}Time evolution of wealth distribution for additive exchange. Different symbols show the wealth distribution at different time steps. A Boltzmann-Gibbs distribution is apparent at $399500$ (MCS). In the inset we can see the fit to an exponential function.}
\end{figure}

\begin{figure}[!htp]
 \centering
 \includegraphics[scale=0.3, angle=0]{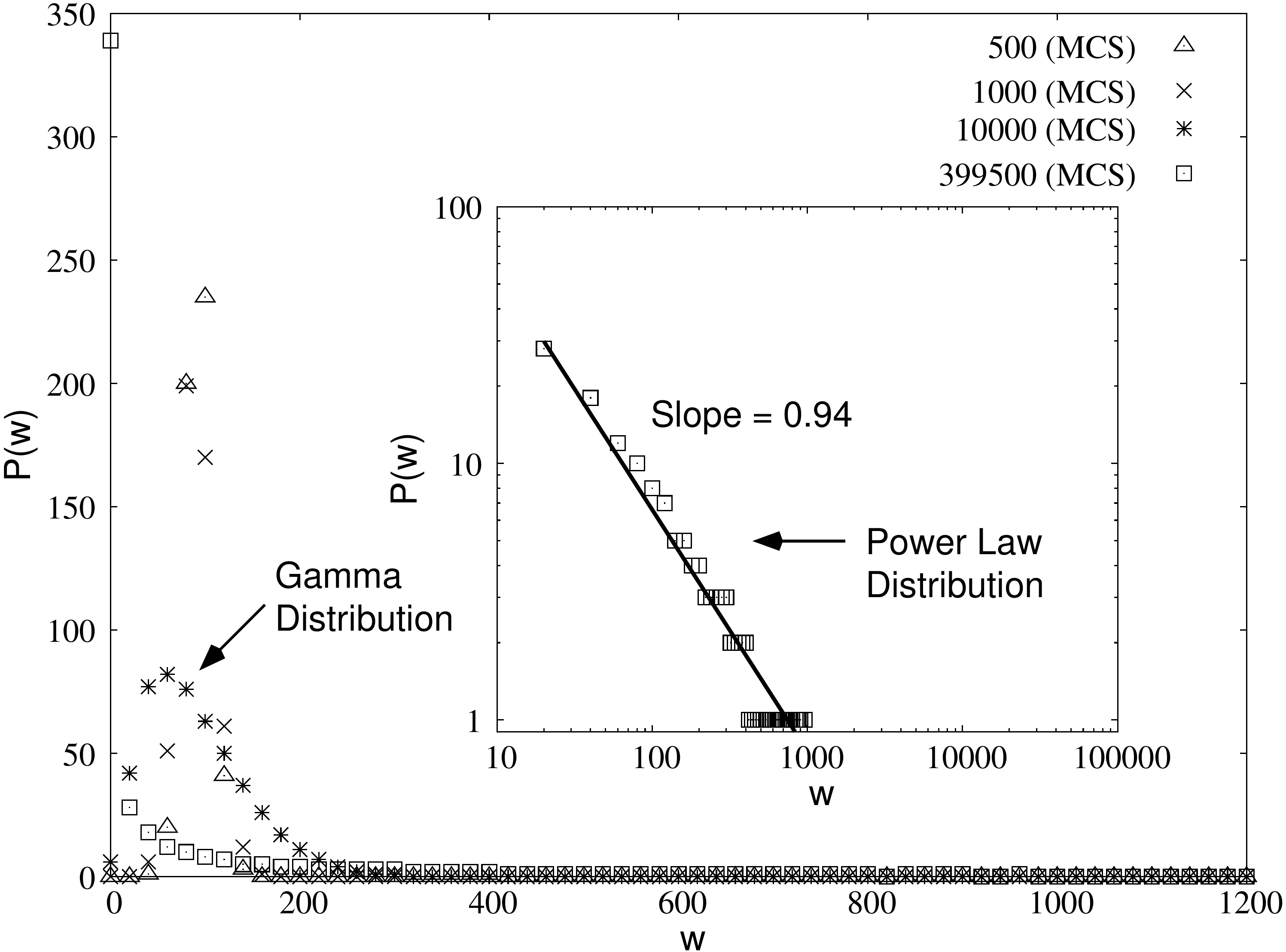}
 \caption{\label{fig:1b} Same as figure \ref{fig:1a} but for the case of multiplicative exchange. Here we get a power law distribution at $399500$ (MCS). The inset shows the fit to a power law.}
\end{figure}

\noindent The time evolution of the wealth distribution for both the 
additive and multiplicative exchange cases has been studied. 
In Figure \ref{fig:1a} we display the results, which are not normalized, 
for the additive case, where the system evolves to a stationary state and 
the distribution is of the \emph{Boltzmann-Gibbs} type (remark inside). 
These results are already reported \cite{art:12,art:4}. In Figure 
\ref{fig:1b}, for the multiplicative case, we can see that the system 
undergoes different phases including the \emph{Gamma} distribution  and 
the \emph{power law} (remark inside). Again these results are already 
reported \cite{book:9,book:10,art:7}. 
For a sufficiently long time, as already noted in previous 
works \cite{art:11}, the system ends in a condensed state, 
where $N-1$ agents belong to the  poverty state, which is defined as the minimum state of wealth where the amount of money is less than $\ddinero$, and a single
agent who is in the maximum state of wealth. 

\begin{figure}[!htp]
 \centering
 \includegraphics[scale=0.3, angle=0]{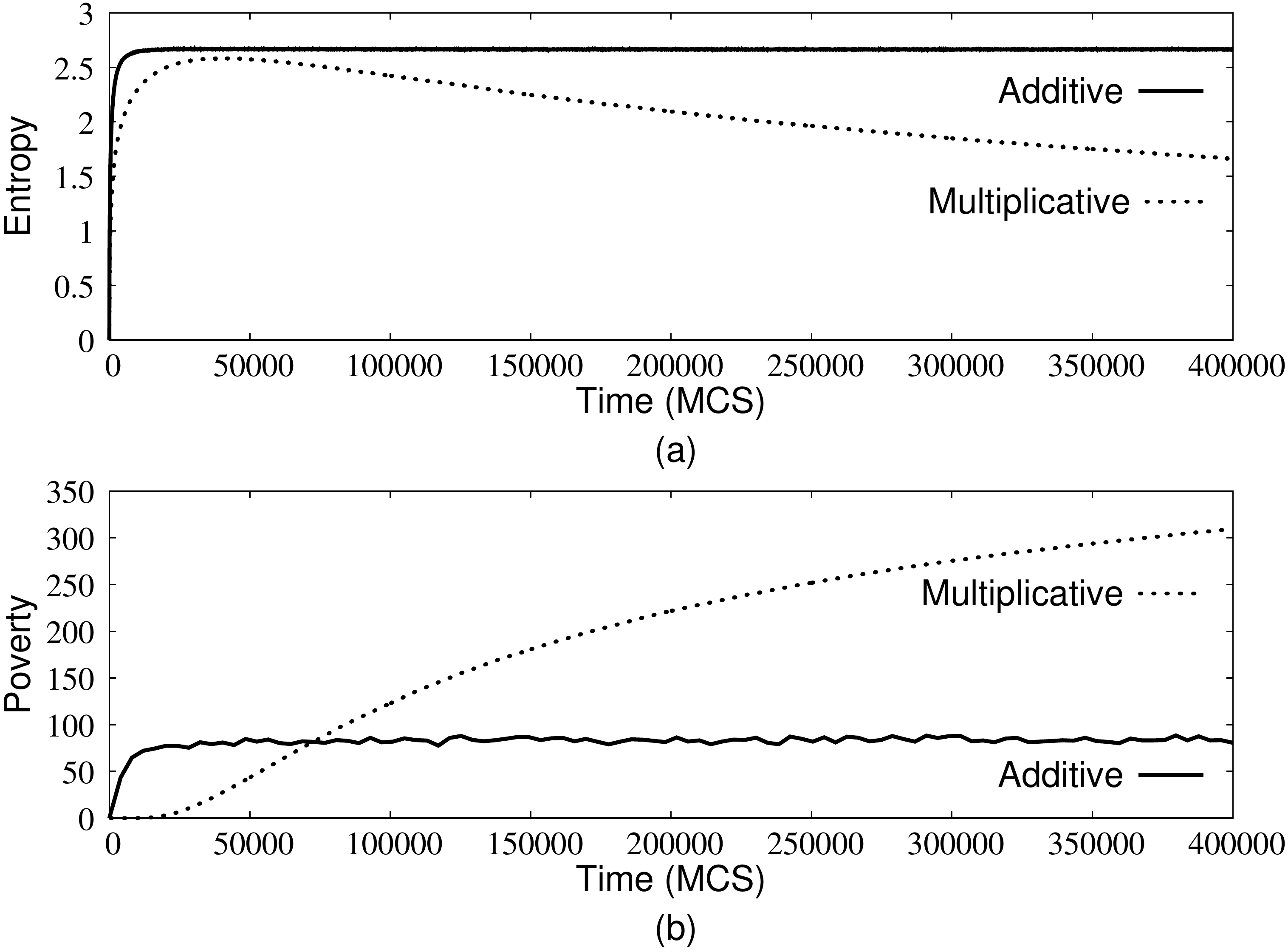}
 \caption{\label{fig:2}Time evolution of entropy (a) and poverty (b) for both cases of exchange, additive and multiplicative, in a log-normal scale.}
\end{figure}

\noindent We have also studied the Shannon entropy, defined by
\begin{equation}
 S = - \sum_{k} P_k \ln P_k ,\label{eq:5}
\end{equation}
where $P_k = n_k/N$, that is, the probability for the agent
to be in the state $k$, with wealth $w_k$.
In Figure \ref{fig:2}(a) and \ref{fig:2}(b) we display the behavior of the 
Shannon entropy as well of the poverty for the additive and multiplicative 
exchanges. In the multiplicative case the entropy reaches a maximum 
before decreasing monotonically, therefore the number of states 
available to the system is minimum at the condensed state, $S_c$
\footnote{The entropy of condensation $S_c$ is calculated as:
\[S_c = -\left( \frac{N-1}{N} \ln \frac{N-1}{N} + \frac{1}{N} \ln \frac{1}{N}\right)\] }. 
In our case, with the parameters given for the system, this value is 
$S_c=0.01442$

\subsection{Exchange with bankruptcy}
\noindent We define that an agent is bankrupt, in the additive case, when it ends up with no money to make more bets ($w < \ddinero$). In the multiplicative case, an agent is bankrupt when the closest integer of $\nu\cdot w$ is zero. Hence if the agent is bankrupt, it is not allowed to participate in the
dynamics anymore.  As expected, the condensation point is reached 
faster than in the previous cases when bankruptcy was not included. In those 
cases, the agents who have reached the absolute poverty were still 
allowed to participate in the game, and in principle they could become rich, 
even though this situation is very unlikely.
\begin{figure}[!htp]
 \centering
 \includegraphics[scale=0.3, angle=0]{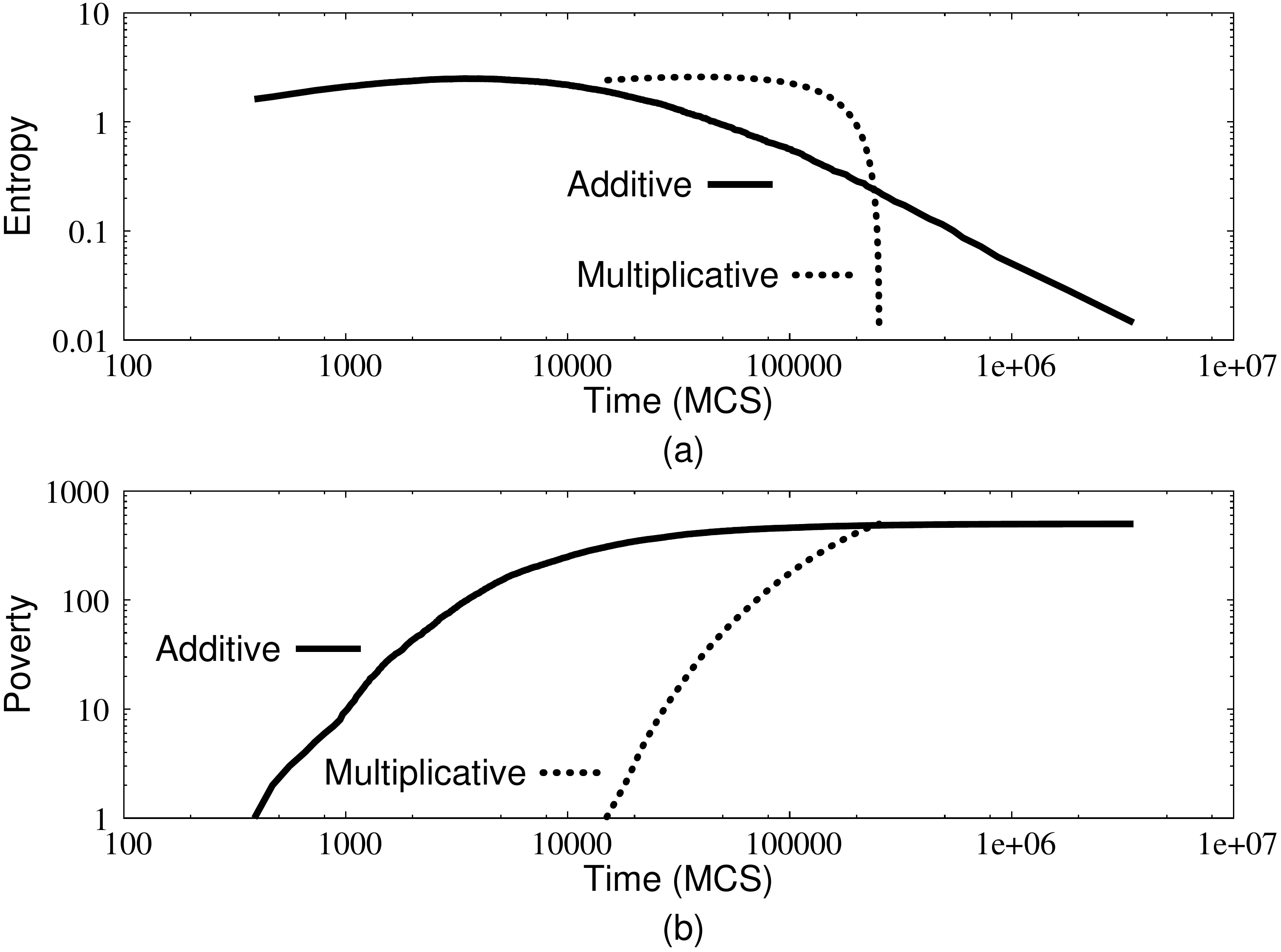}
 \caption{\label{fig:3}Time evolution of entropy (a) and poverty (b) for both cases of exchange, additive and multiplicative, using the bankruptcy concept, in a log-log  scale.}
\end{figure}

\begin{figure}[!htp]
 \centering
 \includegraphics[scale=0.3, angle=0]{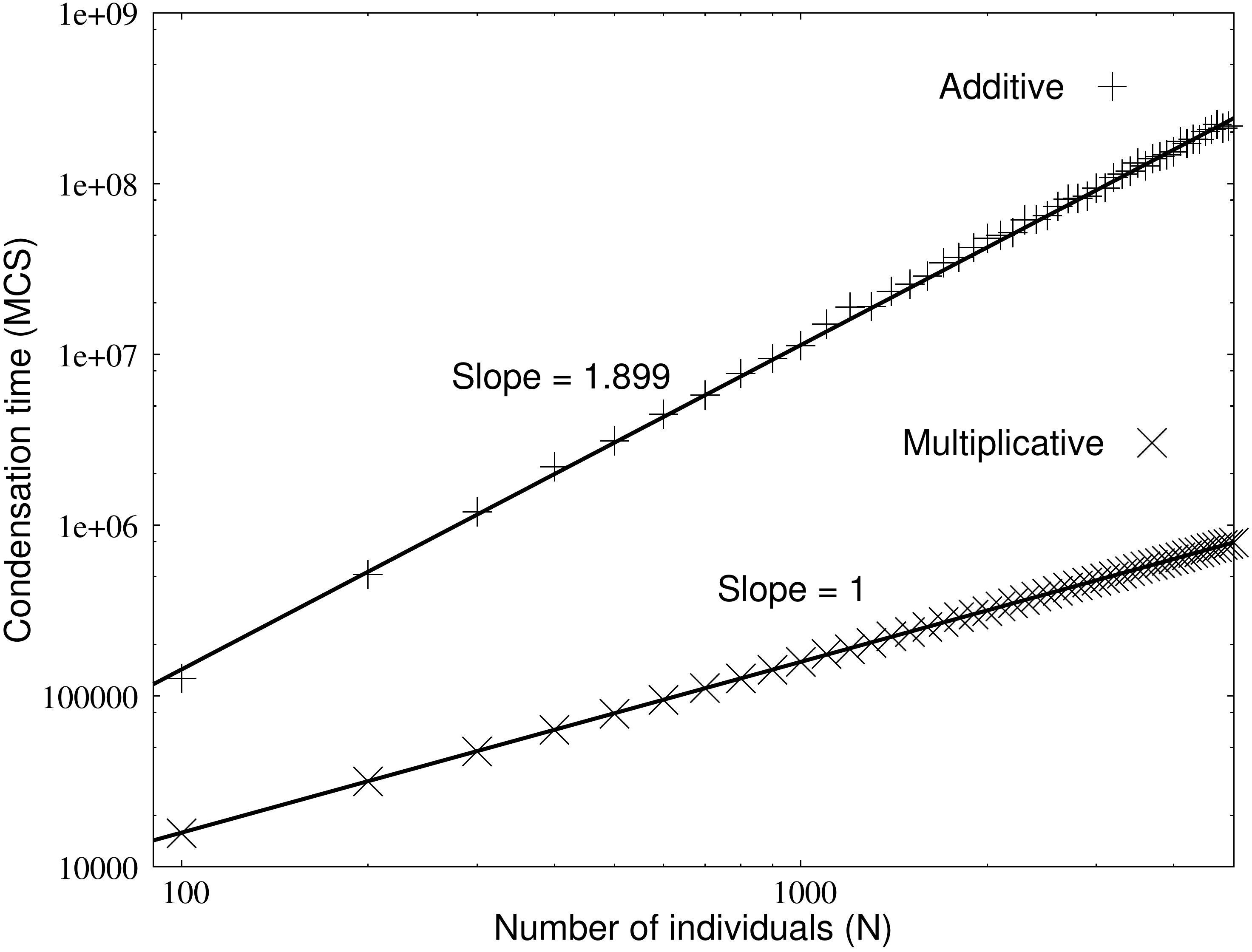}
 \caption{\label{fig:4}Time of condensation vs number of agents for both cases of exchange, additive and multiplicative using the bankruptcy concept, in a log-log scale. Note that the relation between time of condensation and number of individuals is different for each case of wealth exchange.}
\end{figure}

\noindent Figure \ref{fig:3} shows the results with bankruptcy.
It is important to notice that in both cases, additive and multiplicative, 
the system reaches the condensed phase. Poverty reaches its maximum value, 
$N-1$ or $499$, and entropy ends up at the $S_c$ value previously given. 
Since in both cases we 
obtain condensation, we study the time it takes to reach this point 
when we vary parameters such as $N$, the number of agents, and $\ddinero$, 
the fraction of the wealth exchange. In Figure \ref{fig:4} we show this 
variation with respect to $N$ and found a different behavior depending 
on the exchange case. In the additive case the dependence of the 
condensation time 
with respect to $N$ goes like $t_c^{add} \propto N^{1.899}$ whereas in the 
multiplicative case we get a linear dependence $N$ $t_c^{mul} \propto N$. 

\begin{figure}[!htp]
 \centering
 \includegraphics[scale=0.3, angle=0]{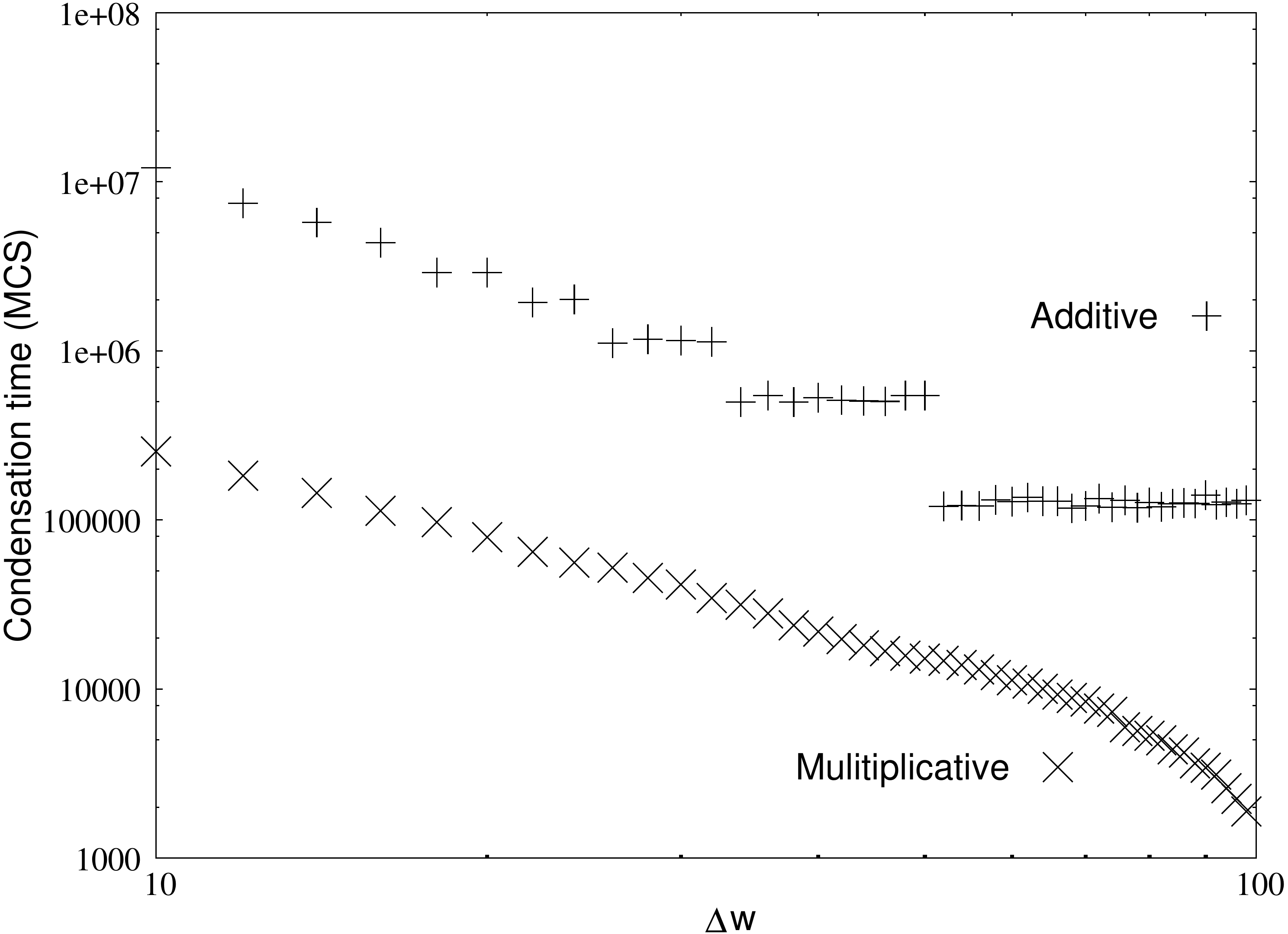}
 \caption{\label{fig:5}Time of condensation vs $\ddinero$ for both cases of exchange, additive and multiplicative using bankruptcy concept, in a log-log scale. Again the relation between time of condensation and $\ddinero$ is different depending on the exchange case.}
\end{figure}

\noindent The dependence of the condensation time with $\ddinero$ is 
depicted in Figure \ref{fig:5}. In the additive case 
there is a discrete behavior due to the fact that the condensation time has 
a well defined value for the different ranges of $\ddinero$. This can be 
explained if we consider the condensation time as being proportional to 
the minimum number of exchanges $min_{ex}$ that two agents can 
make before one of them reaches poverty. Hence we write $min_{ex}$ as
\[min_{ex}=\pentera{\frac{\promedio{w}}{\ddinero}}\text{.}\] 
For instance if we considered $\promedio{w} = 100$ and $\ddinero = 50$, 
two agents ($i$,$j$) will have a minimum number of exchanges $min_{ex} = 2$, that 
means that there are 2 exchanges at least, before one agent reaches poverty. 
If we have $\promedio{w} = 100$ and $\ddinero = 51$, the minimum number of exchanges 
is now $min_{ex} = 1$, and only one bet can be made. Therefore the 
time of condensation depends of the minimum of exchange.

\noindent In a multiplicative exchange
the behavior of the condensation time $t_c^{mul}$ do not shows the discrete situation of the 
additive case since the bet is not constant. One would expect that if we 
increase the amount of the wealth exchange in each transaction, the time it 
takes to condensate gets lower, and that we can see in Figure \ref{fig:5}.

\section{Dynamics of the exchange in a social network}

\noindent In this section we consider a system composed of $N$ agents 
placed in an undirected non-fully connected network with a degree 
distribution given by equation (\ref{eq:7}). It is clear that the agents 
cannot interact freely, but instead they do it through relations that form 
a social network. One would expect that the dynamics shows some 
differences with respect to the urn case.

\noindent Since the number of links varies among the agents, we will 
consider the average value, or connectivity $\promedio{k}$ 
(see section \ref{sec:2}), for a given network. We consider again the 
additive and multiplicative cases, without taking bankruptcy into account since, as we have seen, its main effect is to drive the system into the condensed state. Therefore if we introduce it in the undirected non-fully connected network, we are going to have local condensations, due to the presence of several components, obtaining as a result wealth distributions similar to multiplicative exchange which is going to study in section \ref{sec:unfcn-me}. 

\subsection{Additive exchange}

\begin{figure}[!htp]
 \centering
 \includegraphics[scale=0.3, angle=0]{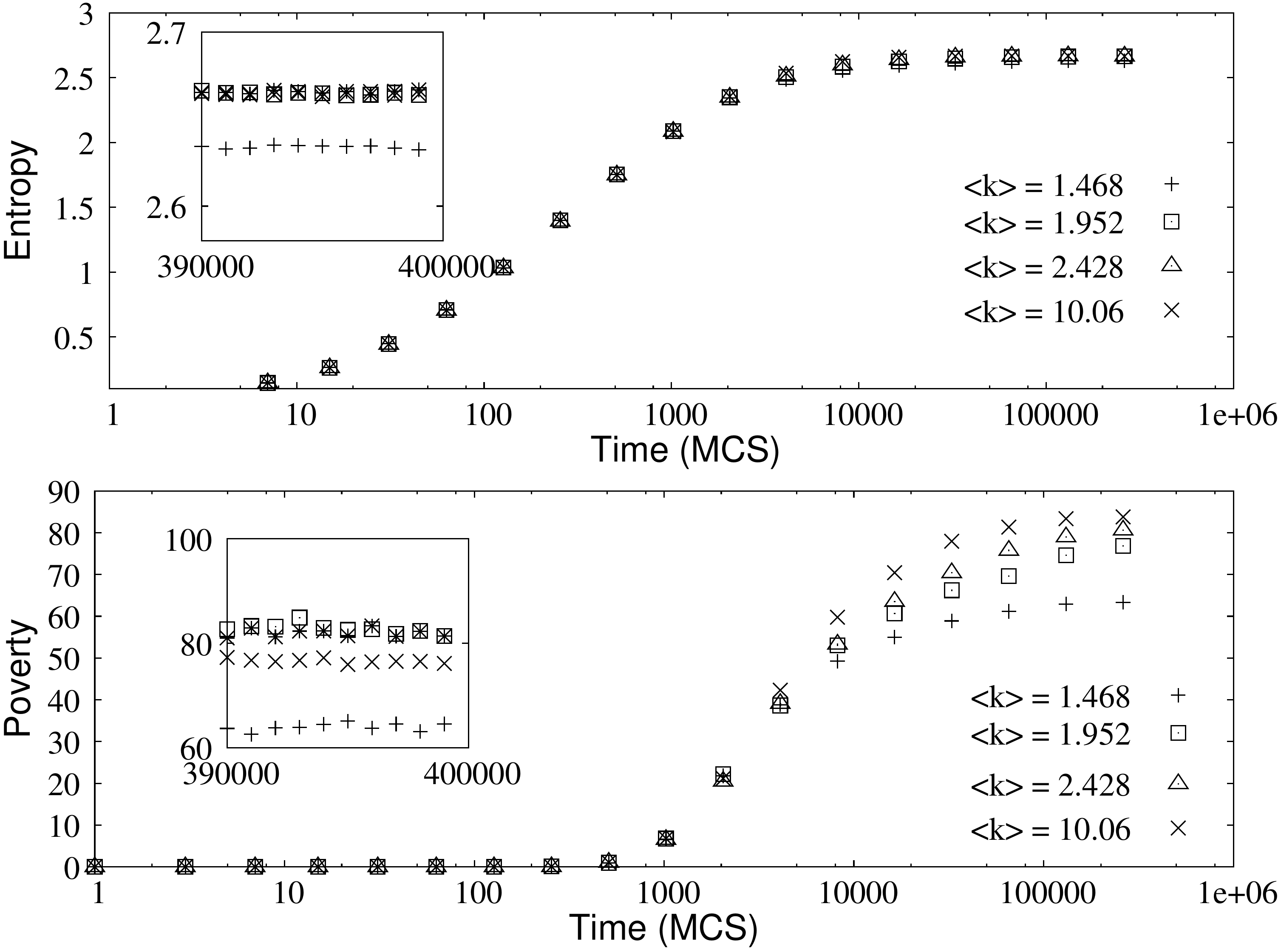}
 \caption{\label{fig:7}Time evolution of entropy (a) and poverty (b) in the additive exchange case for different values of $\promedio{k}$. In the inset we appreciate that for $\promedio{k}=1.468$ the asymptotic entropy is the lowest. The same behavior occur to the poverty value.}
\end{figure}

\noindent In Figure \ref{fig:7} we display the temporal evolution of entropy 
and poverty for the additive case and for different values of $\promedio{k}$. 
The temporal evolution of entropy is very similar in all cases, 
and furthermore, always lead to a stationary distribution of wealth, 
no matter the average connectivity value $\promedio{k}$. The entropy, 
as well as the poverty, reaches a maximum value. An interesting 
thing to notice is that the connectivity value of $1.468$ leads to
lower values of entropy and poverty than those with other connectivities 
(see insets in Figure \ref{fig:7}). In order to understand this, we will
examine the component size distribution $\chi(s)$, which gives the probability to find a component with size $s$ in the network.  

\begin{figure}[!htp]
\centering
\includegraphics[scale=0.3, angle=0]{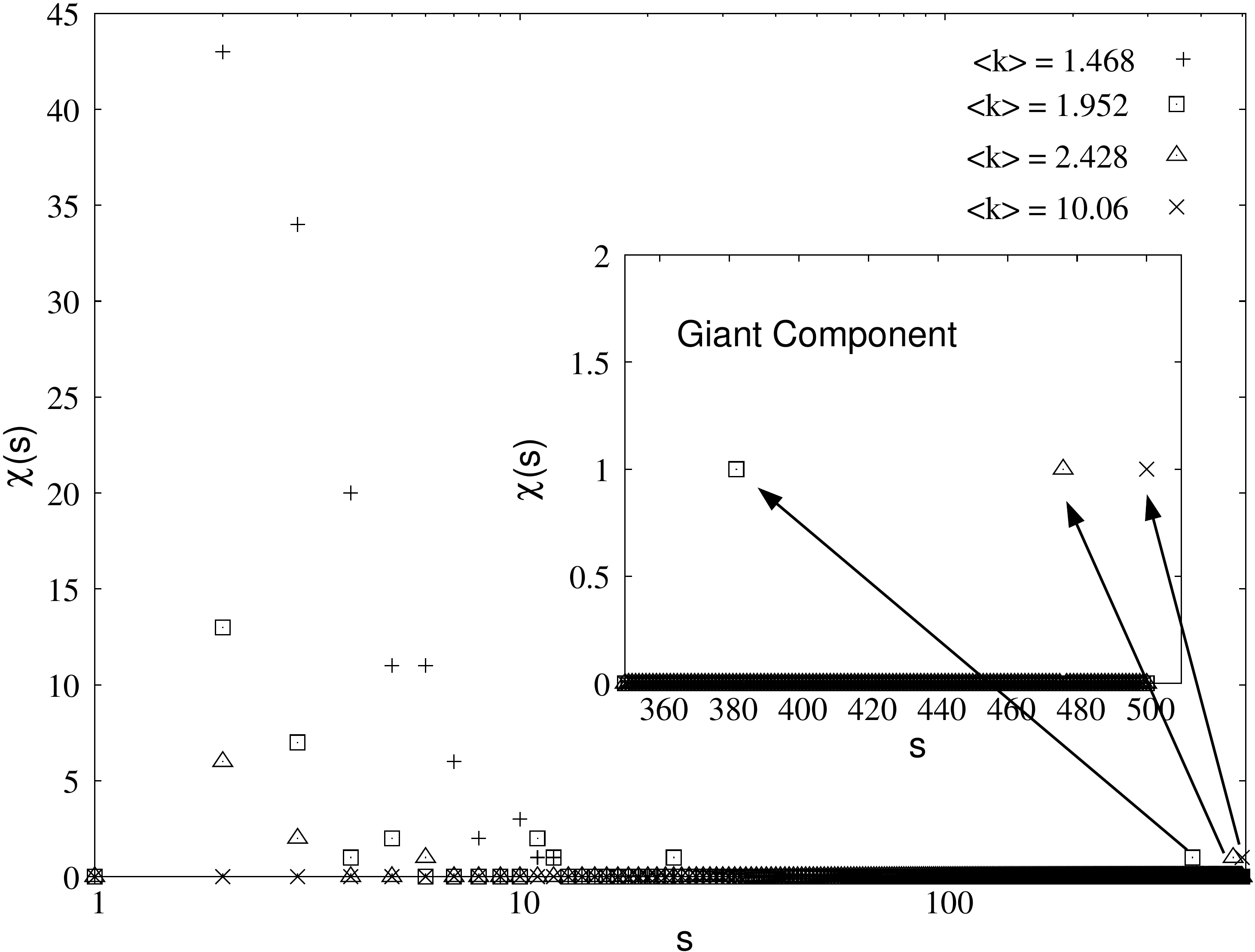}
\caption{\label{fig:12}Component size distribution in the network for different values of connectivity $\promedio{k}$. One can see that for $\promedio{k} = 1.468$ there is no giant component, as opposed to the other three cases (see inset).}
\end{figure}

\noindent In Figure \ref{fig:12} we display the component size distribution in the network, for each average connectivity value $\promedio{k}$. The distribution is not normalized since this makes no difference to the discussion. For $\promedio{k} = 1.468$, we observe that components with sizes of $2$, $3$ and $4$ are predominant in network. Therefore in the network are present groups of linked agents that are isolated from the rest of the network.
The exchange becomes isolated from the rest of the system, and 
the effect of this situation is to produce economic classes as the system evolves. This effect is more easily obtained when $\promedio{k}$ is lower, as one would expect. Therefore, the isolated 
agents that form the component are decoupled from the system's dynamics, and they never reach absolute 
poverty. The more components we get the less poverty we obtain.

\begin{figure}[!htp]
 \centering
 \includegraphics[scale=0.3, angle=0]{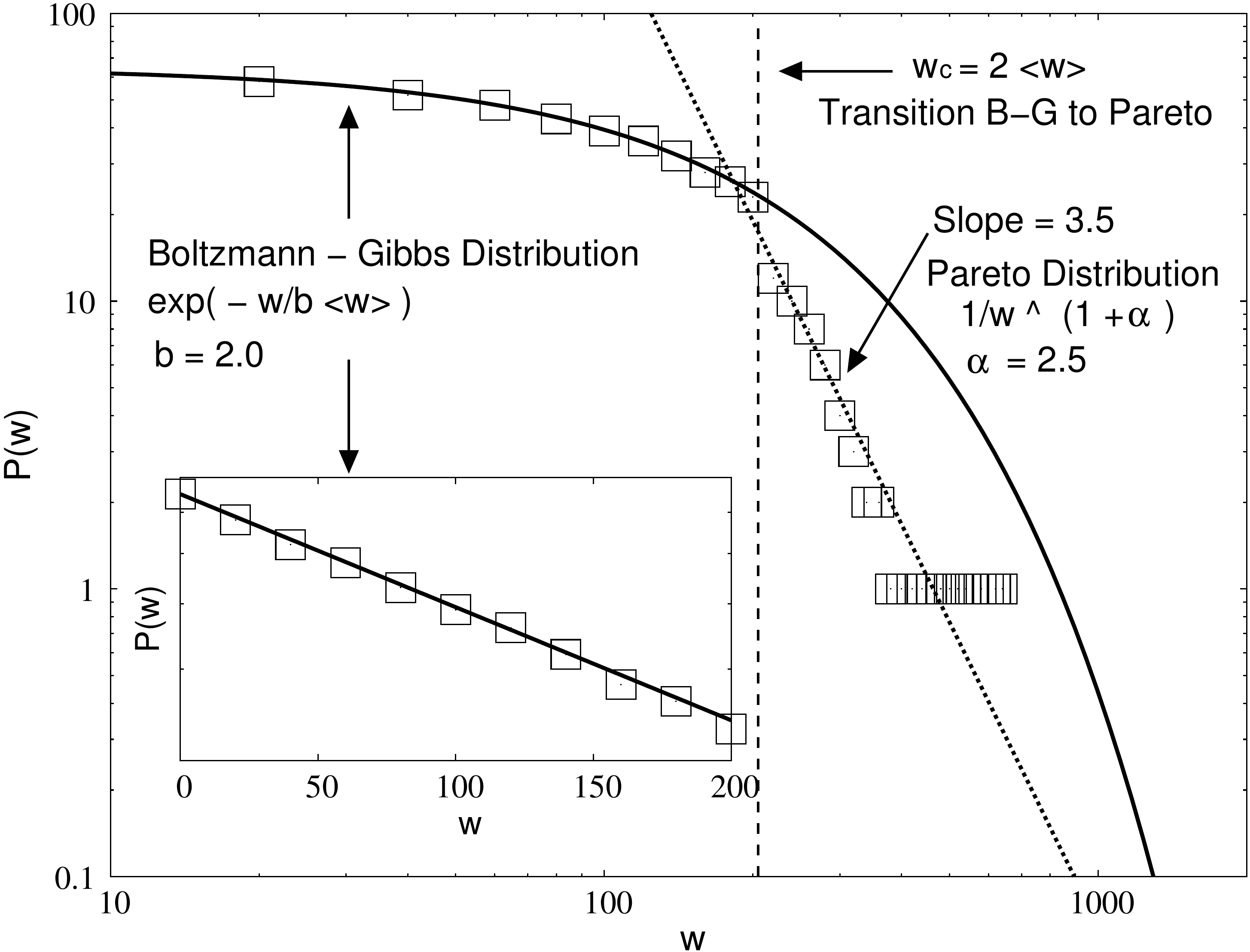}
 \caption{\label{fig:8a}Wealth distribution in the additive exchange case, in time step $399000$ MCS for  $\promedio{k} = 1.468$. The solid curve fits an exponential function (see inset for a log-normal scale), while the dashed curve fits to a power law function. Two different regimes are clearly seen separated at $2\promedio{w}$: a Boltzmann-Gibbs distribution for values below to $2\promedio{w}$, and a Pareto distribution for those above to $2\promedio{w}$. This behavior is similar to Figure \ref{fig:intro}.}
\end{figure}

\begin{figure}[!htp]
 \centering
 \includegraphics[scale=0.3, angle=0]{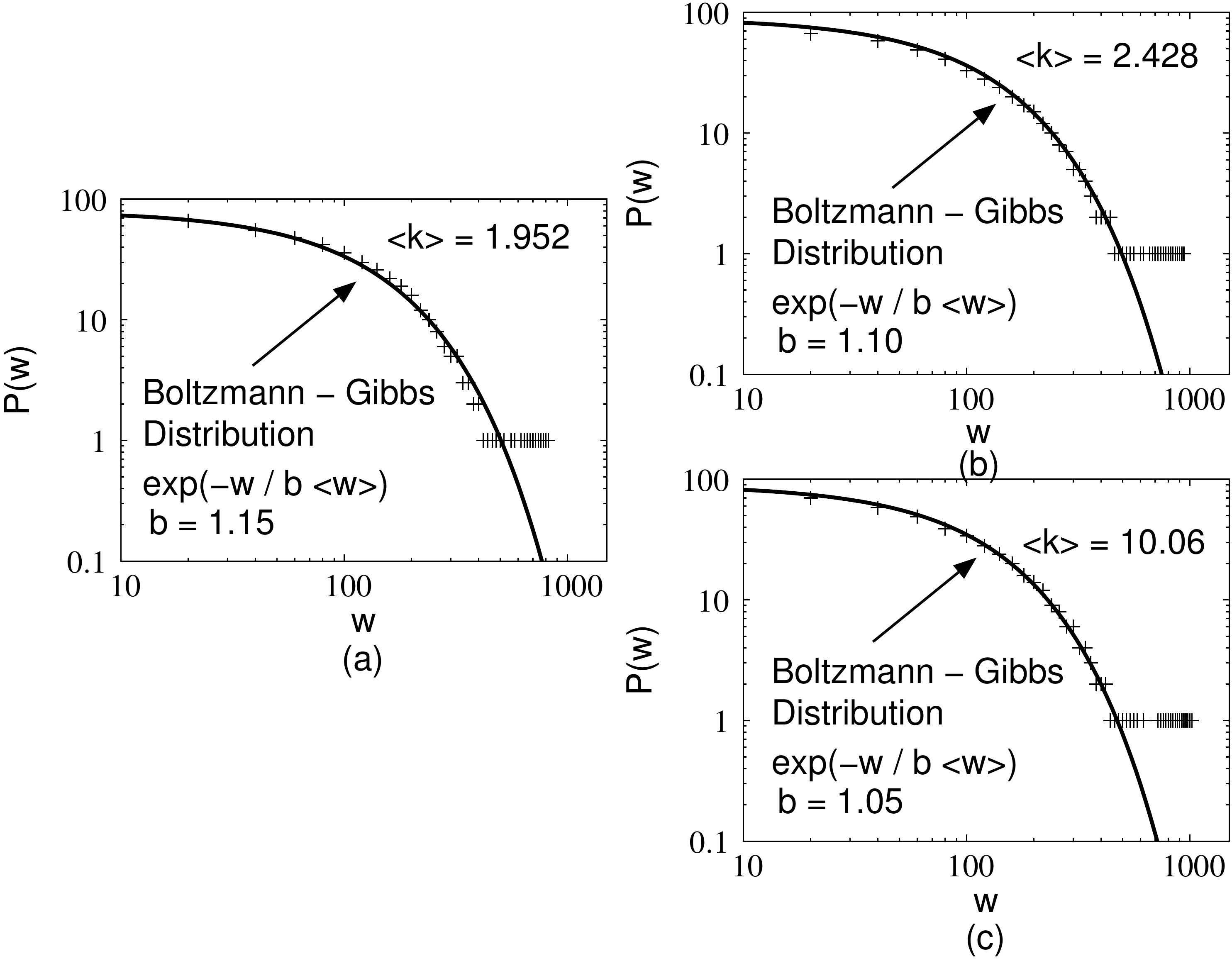}
 \caption{\label{fig:8b}Wealth distribution in the additive exchange case, at time step $399000$ MCS for values of $\promedio{k}$ other than $1.468$. The data for each value of $\promedio{k}$ can be fitted to the exponential function adjusting the value of parameter $b$ (see table \ref{tab:1}).}
\end{figure}

\noindent Figure \ref{fig:8a} shows the wealth distribution for additive exchange at time step ($t=399000$ MCS), for the lower value of $\promedio{k} = 1.468$. We notice two social classes, one where the wealth of the agents is below $2\promedio{w}$ and with a \emph{Boltzmann - Gibbs} distribution (continuous line Figure \ref{fig:8a}) of the form

\begin{equation}
 P(w) = Ce^{-\frac{w}{b\promedio{w}}}, \label{eq:3}
\end{equation}
where $b$ is equal to $2$. The other class, with wealth above $2\promedio{w}$ behaves like a Pareto distribution (dashed line) of the form

\begin{equation}
 P(w) = \frac{C}{x^{1 + \alpha}}, \label{eq:4}
\end{equation}
with $\alpha =  2.5$. This particular behavior, the coexistence of two different wealth distributions, has been observed in real economic data (see Figure \ref{fig:intro}). It is important to mention that these two classes 
coexist, and these two distributions are joint at a wealth value of 
$w=2\promedio{w}$. This value arises due to the large number of components
of size of $2$ in the network, allowing to each agent obtain a maximum 
wealth of $2\promedio{w}$. 
 
\noindent Other authors \cite{art:17,art:16} had found both distributions in
coexistence,  but using a more sophisticated model, and with a different 
network topology. 
 
\noindent This formation of two classes arises from the fact
that at a value of $\promedio{k}=1.468$, most of the components are formed by two agents (see Figure \ref{fig:12}).
Therefore, the poor sector of the system is formed mostly by groups of 2 agents, while the rich sector have to the possibility to be formed by components of 3 or more agents, which allows an agent to have a wealth higher than $2\promedio{w}$.

\noindent When $\promedio{k}$ increase, the components formed by two agents start to decrease (see Figure \ref{fig:12} around $s=2$), and as a consequence, the two wealth distributions observed in Figure \ref{fig:8a} disappear. We end up only with a \emph{Boltzmann - Gibbs} distribution for each $\promedio{k}$ value, as we can see in Figure \ref{fig:8b}. These \emph{B-G} distributions of Figure \ref{fig:8b} are obtain adjusting a parameter $b$ (see eq \ref{eq:3}) for each value $\promedio{k}$. This is shown in table \ref{tab:1}.
\begin{table}[!h]
 \begin{center}
 \begin{ruledtabular}
  \begin{tabular}{cccc}
   &$\promedio{k}$&$b$&\\
   \hline
   &$1.952$&$1.15$&\\
   &$2.428$&$1.1$& \\
   &$10.06$&$1.05$&\\
  \end{tabular}
 \end{ruledtabular}
\caption{\label{tab:1}Values of parameter $b$ (eq \ref{eq:3}) for different average connectivities $\promedio{k}$ in the additive exchange case.}
 \end{center}
\end{table}

\noindent Note that the systems reach the equilibrium at temperatures higher than temperature in the undirected fully connected network case. That means the value of $b$ is greater than $1$, and therefore the product $b\promedio{w}$ (that can be considered like a temperature) is higher than $\promedio{w}$ (equilibrium temperature in an undirected fully connected network). Also, as $\promedio{k}$ increases, the value of $b$ tends to 1, which makes 
sense since a very large value of $\promedio{k} \to N-1$ goes to an undirected 
fully connected network.

\noindent It is important to remark that the formation of the components in the 
network allows the exchange transactions to occur in a local way, and that 
prevents a population of agents to become poor. The agents are in 
a \emph{Boltzmann - Gibbs} distribution, since in the component the maximum 
possible wealth is $2\promedio{w}$. Therefore these components prevent 
agents from becoming poor, but also rich.

\subsection{Multiplicative exchange}\label{sec:unfcn-me}

\begin{figure}[!htp]
 \centering
 \includegraphics[scale=0.3, angle=0]{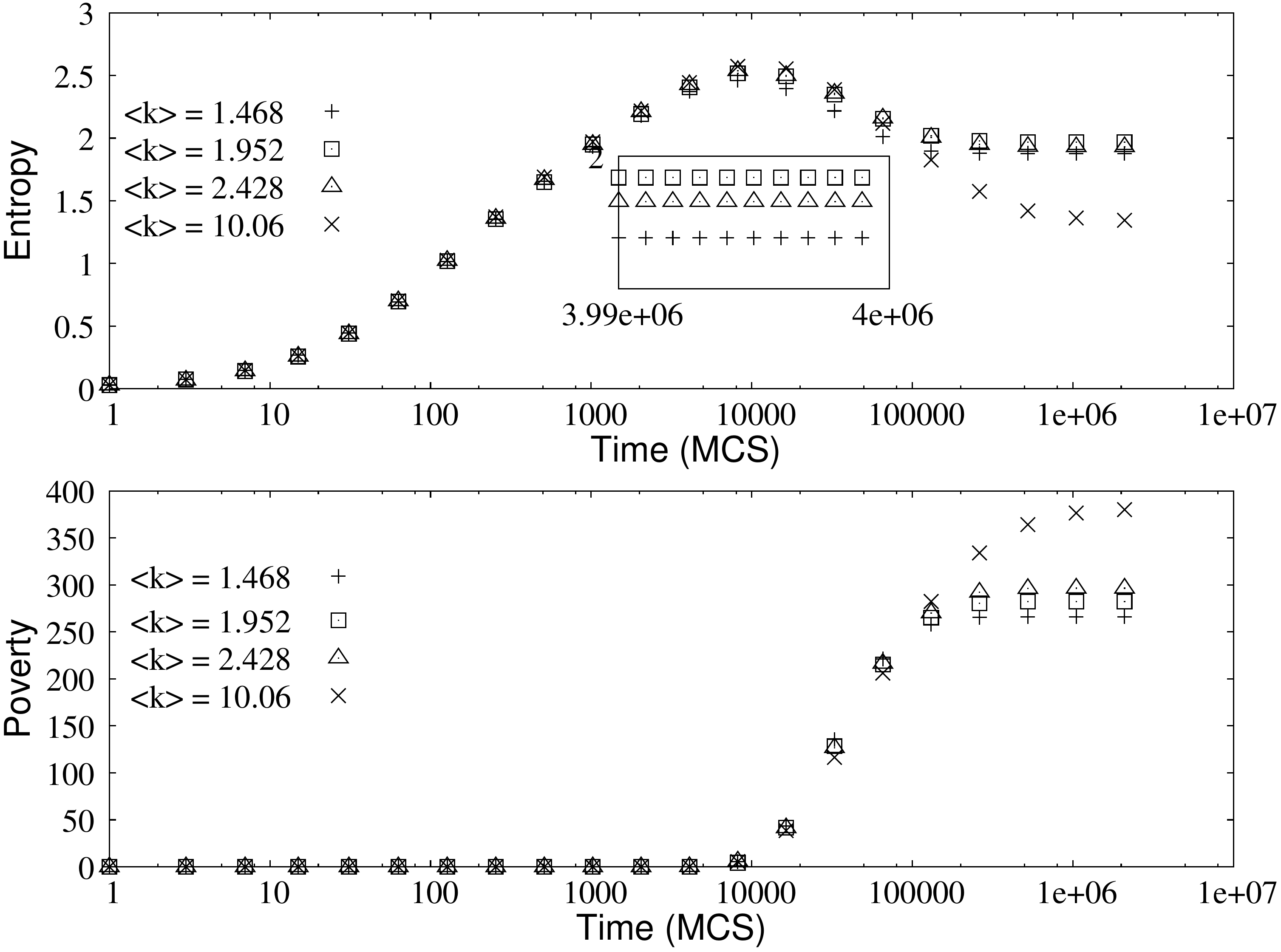}
 \caption{\label{fig:10}Time evolution of entropy (a) and poverty (b) in the multiplicative exchange case for different values of $\promedio{k}$. Note that, when there is a giant component, the entropy and poverty are inversely proportional ($\promedio{k}\neq 1.468$). In the absence of a giant component, that is not the case.}
\end{figure}

\noindent In Figure \ref{fig:10} we show the temporal evolution of 
entropy and poverty for the multiplicative case.  One can see that entropy 
goes through a maximum before decreasing and reaching a constant value 
that is different for each value of $\promedio{k}$. The system with 
larger parameter $\promedio{k} = 10.06$ is the one with more possibilities 
to condense, as in the undirected fully connected network case, because this has a component size 
of $500$. We can also notice that its entropy is the lowest and its 
poverty the highest. However we expect that condensation occurs later than 
in the undirected fully connected network case.

\begin{figure}[!htp]
 \centering
 \includegraphics[scale=0.3, angle=0]{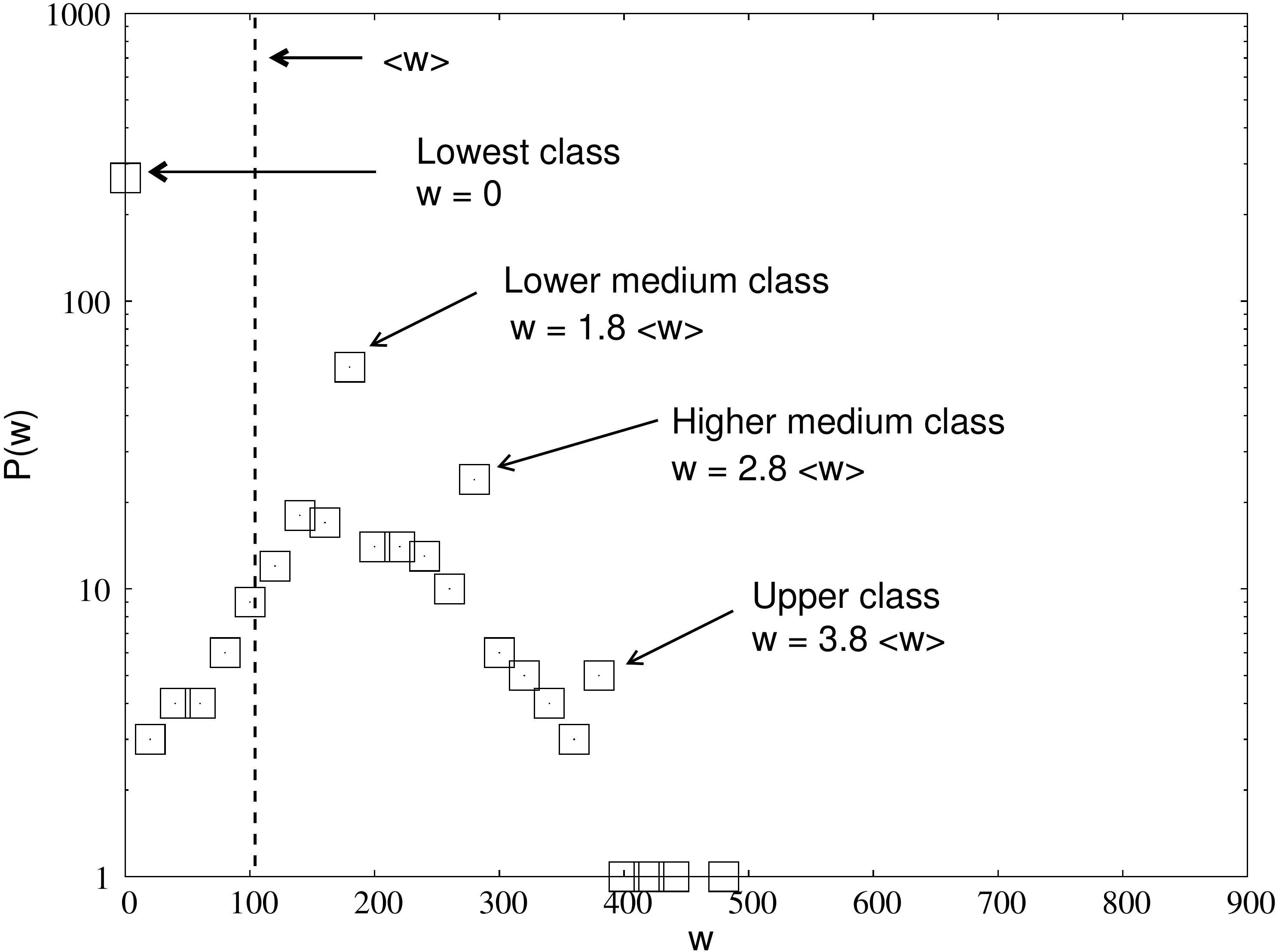}
 \caption{\label{fig:11a}Wealth distribution in the multiplicative exchange case, at time step $399000$ MCS for an average connectivity value of $\promedio{k}=1.468$. We can see in this case the presence of various well defined wealth classes, that appear because there is no giant component.}
\end{figure}

\begin{figure}[!htp]
 \centering
 \includegraphics[scale=0.3, angle=0]{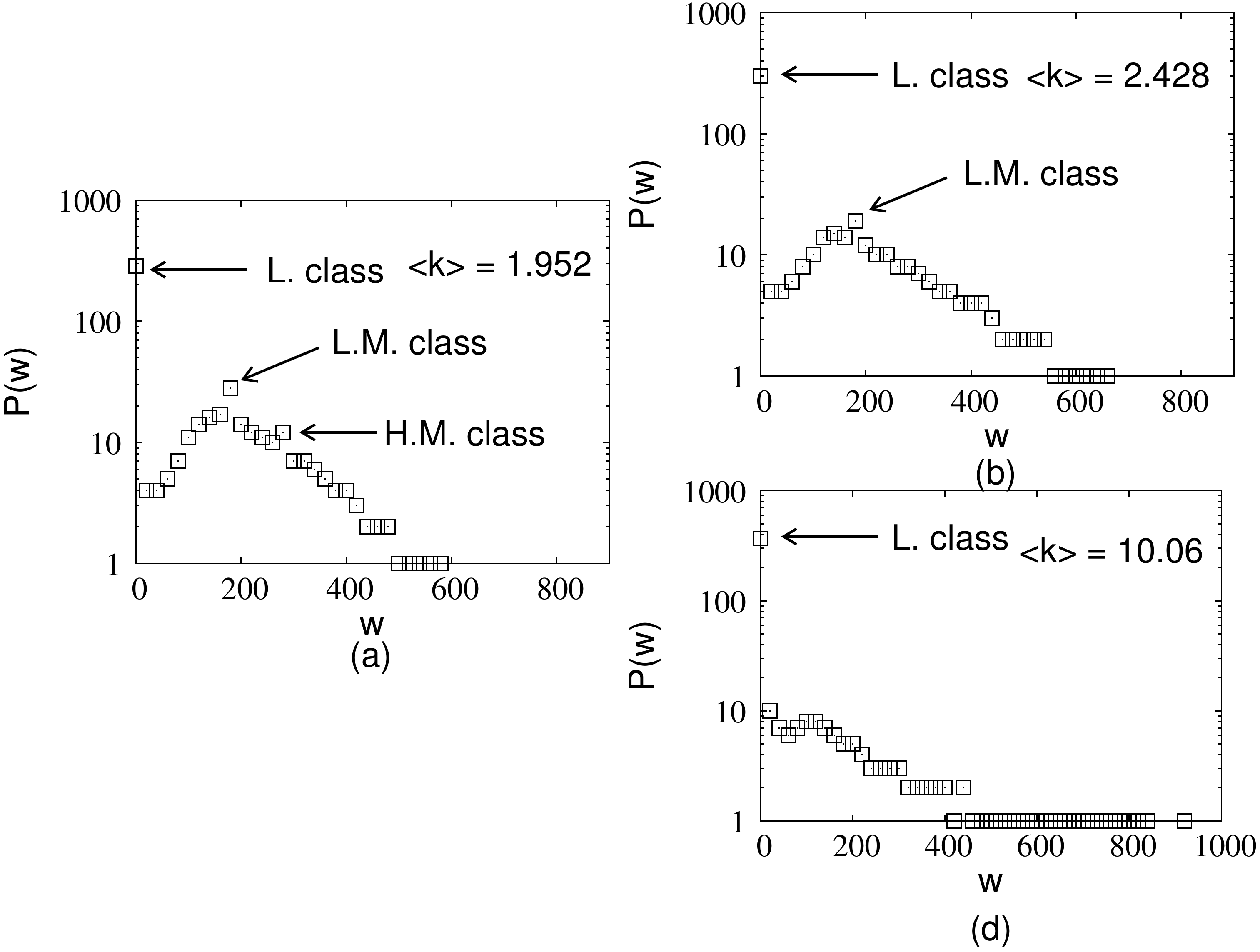}
 \caption{\label{fig:11b}Wealth distribution in the multiplicative exchange case, at time step $399000$ MCS for values of $\promedio{k}$ different to $1.468$. We observe that as $\promedio{k}$ increase, the classes tend to disappear. At the end a bimodal behavior is obtained.}
\end{figure}

\noindent Figure \ref{fig:11a} shows the  wealth distribution for a 
multiplicative exchange, in time step ($t=399000$ MCS) for the lower value 
of $\promedio{k}=1.468$. A very interesting phenomenon  can be observed. The 
appearance of four well defined classes with wealth of zero, 
$1.8\promedio{w}$, $2.8\promedio{w}$ and $3.8\promedio{w}$ respectively 
(isolated points in the figure). These classes of wealth will eventually 
become 0, $2\promedio{w}$, $3\promedio{w}$ and $4\promedio{w}$ when the 
system reaches the stationary state at later times.
The isolated point of wealth $1.8\promedio{w}$ corresponds to the presence 
of groups of two agents in the system, because, if two agents interact only 
with each other, the maximum value of wealth that a agent will have 
is $2\promedio{w}$. Similarly, the points with wealths of $2.8\promedio{w}$ 
and $3.8\promedio{w}$ correspond to the presence of closed groups of 
three and four agents respectively. 
This is in agreement with Figure \ref{fig:12}, where the most prevailing 
component sizes are those of values $2$, $3$ and $4$ respectively. 
In addition to these classes, we observe a background of agents with 
variable wealths. These background arises from agents that belong to 
components of sizes larger than $4$. In this case there is a larger flux 
of money among many agents in the network. 

\noindent It is clear that classes appear due to the formation of components with different sizes. A 
local condensation occurs in each one, and this gives us a hint about the 
way the system behaves when the value $\promedio{k}$ decreases: the entropy 
of the system should reach a minimum which is stationary and in this point 
half of the agents are poor and the other half has a wealth 
of $2\promedio{w}$. As soon as $\promedio{k}$ starts to  
increase, the separation between the classes begins to disappear, 
until we are left with a class of agents with wealth 0, and another class 
on non-zero wealth agents, as seen in Figure (\ref{fig:11b}a - \ref{fig:11b}c).
A Similar bimodal behavior has been obtained in Ref \cite{art:6}.

\subsection{Class formation and transition to the giant component}
\noindent From previous discussions, we observed an interesting 
phenomenon (appearance of wealth classes) in the case of an undirected 
non-fully connected network, which has a maximum  connectivity equal to $2$. 
We shall see in more detail the transition to the giant component. Starting 
from generating function proposed in \cite{art:19}, but considering that
in our case the sum is finite and begins in $k=1$, we can write 
\begin{equation}\label{eq:8}
G_0(x) = \sum_{k=1}^{k_{max}} p(k)x^k.
\end{equation}
\noindent This function allows us to know if a giant component is present 
in the network. The average number of links, or connectivity mean 
value $\promedio{k}$, is given by 
\begin{equation}\label{eq:9}
z_1 = \promedio{k} = \sum_{k=1}^{k_{max}} kp(k) = G_0'(1),
\end{equation}
\noindent thus the average number $z_2$ of second neighbors will be 
\begin{equation}\label{eq:10}
z_2 = G_0''(1).
\end{equation}
\noindent In the network there is a giant component if the condition $z_2 > z_1$ is satisfied.

\noindent Using the value of $p(k)$ from equation (\ref{eq:7}), we can 
write $G_0(x)$ as
\begin{equation}\label{eq:11}
G_0(x) = \frac{x}{k_{max}}\left[\frac{1-x^{k_{max}}}{1-x}\right].
\end{equation}

\noindent Now we can calculate the values of $z_1$ and $z_2$ for each value 
of $k_{max}$. These values are shown in table \ref{tab:2}.
\begin{table}[!h]
 \begin{center}
  \begin{ruledtabular}
   \begin{tabular}{ccccc}
     & $k_{max}$ & $z_1$ & $z_2$ &\\
     \hline
     & $2$ & $1.5$ & $1$ &\\
     & $3$ & $2.0$ & $2.66$ &\\
     & $4$ & $2.5$ & $5.0$ & \\
     & $20$ & $10.5$ & $133$ &
   \end{tabular}
  \end{ruledtabular}
  \caption{\label{tab:2}Values of $z_1$ and $z_2$ for each value of $k_{max}$ used in this work. We can see that only for $k_{max} = 2$, we get $z_1 > z_2$, 
which means there is no giant component present.}
 \end{center}
\end{table}

\noindent As we can see, the only value of $k_{max}$ that does not led  to 
a giant component is $k_{max} = 2$. This means that in this case, the system 
has more components than for other values of connectivity, mostly with size of $2$. Consequently, 
the wealth flow is local and therefore the classes arise. Another important 
point is the transition to a giant component, which happens from 
$k_{max} = 2$ to $k_{max} = 3$. When the network has a giant component 
the wealth flow is not local as in the case of $k_{max} = 2$, and the 
classes start to disappear. 

\noindent To see that in the case with $k_{max}=2$ there are more components of size $2$ than other sizes, we calculate $P_s$ and $\chi(s)$. $P_s$ is the probability distribution of cluster sizes, which is the probability of an agent to belong to a size $s$ component. $\chi(s)$ is the component size distribution, which gives the probability to find a component with size $s$ in the network. For $P_s$, we using the following equation
\begin{equation}\label{eq:12}
P_s = \frac{1}{s!}\frac{d^sH_0}{dx^s}\Bigg\lvert_{x=0},
\end{equation}
where 
\begin{equation}\label{eq:13}
H_0(x) = xG_0(H_1(x)).
\end{equation}
\noindent $H_1(x)$ is the generating function for the distribution of the sizes of components which can be written as
\begin{equation}\label{eq:14}
H_1(x) = xG_1(H_1(x)),
\end{equation}
with 
\begin{equation}\label{eq:15}
  G_1(x) = \frac{1}{z}G_0'(x), \quad \text{where } z=\promedio{k}.
\end{equation}
\noindent The component size distribution $\chi(s)$ is given by the equation
\begin{equation}\label{eq:16}
\chi(s) = \frac{1}{\chi_{T}}\frac{P_s}{s}, \quad \text{for } s>1.
\end{equation}
\noindent $\chi_T$ is the normalization factor and is given by
\begin{equation}
\chi_{T} = \sum_s \frac{P_s}{s}.
\end{equation}

\noindent  Using equation (\ref{eq:11}), we calculate $G_1(x)$, and obtain
\begin{equation}\label{eq:17}
G_1(x) = \frac{1}{zk_{max}}\left[\frac{1 - 2zx^{k_{max}} + k_{max}x^{k_{max}+1}}{(1 - x)^2}\right].
\end{equation}
\noindent In our case we are taking $k_{max}=2$, and obtain that $H_1(x)$ is
\begin{equation}\label{eq:18}
H_1(x)=\frac{x}{3 - 2x},
\end{equation}
and $H_0(x)$ is
\begin{equation}\label{eq:19}
H_0(x)=\left(\frac{x^2}{2}\right)\frac{3 - x}{(3 - 2x)^2}.
\end{equation}

\noindent With these functions, we calculate for instance the first seven values of $P_s$ and $\chi_{T} \chi(s)$, which are showed in the table \ref{tab:3}
\begin{table}[!h]
 \begin{center}
  \begin{ruledtabular}
    \begin{tabular}{ccccc}
     &s & $P_s$ & $\chi_{T} \chi(s)$ &\\
     \hline
     & $1$ & $0$ & $0$ &\\
     & $2$ & $1/6$ & $1/12$ &\\
     & $3$ & $1/6$ & $1/18$ &\\
     & $4$ & $4/27$ & $1/27$ &\\
     & $5$ & $10/81$ & $2/81$ &\\
     & $6$ & $8/81$ & $4/243$ &\\
     & $7$ & $56/729$ & $8/729$ &
    \end{tabular}
  \end{ruledtabular}
  \caption{\label{tab:3}First seven values of $P_s$ and $\chi_T \chi(s)$, in the case of $k_{max}=2$. We can see that the probability to find components with size of $2$ in the network is highest. This result is agree with the Figure \ref{fig:12}.}
 \end{center}
\end{table}

\noindent From this table we can see that an agent has the same probability of being in a size $2$ or $3$ component, and these probabilities are larger than for other components. In the other hand, the probability to find in the network a component of $2$ agents ($s=2$) is the highest, this agrees with Figure \ref{fig:12}. For example, suppose that $N$, the number of agents, is $36$ and $k_{max} = 2$. From table \ref{tab:3} we know that in the network there will be $3$ components with size $2$, meanwhile there will be $2$ components with size $3$.

\noindent To show how the number of components is decrease when the $k_{max}$ value is increase, we calculate the mean component size, $\promedio{s}$, using the following equation
\begin{equation}\label{eq:20}
\promedio{s} = 1 + \frac{zu^2}{[1 - S][1 - G_1'(u)]}.
\end{equation}
\noindent S is the fraction of network occupied by the giant component and is given by
\begin{equation}\label{eq:21}
S = 1 - G_0(u),
\end{equation}
while $u$ is the smallest non-negative real solution of
\begin{equation}\label{eq:22}
 u = G_1(u),
\end{equation}
\noindent which is a transcendental equation.

\noindent From equation (\ref{eq:17}), we can see that $G_1(u)$ is a polynomial equation of ($k_{max}-1$) degree. Therefore the first intersection between $u$ and this polynomial equation (the intersection is just the solution of equation (\ref{eq:22})), goes to zero if $k_{max}$ increase, and the mean component size, $\promedio{s}$, in equation (\ref{eq:12}) becomes one. It this means that the components in the network are disappearing when $k_{max}$ is increased, just as observed in Figure \ref{fig:12}.

\begin{table}[!h]
 \begin{center}
  \begin{ruledtabular}
   \begin{tabular}{ccccc}
     & $k_{max}$ & $u$ & $\promedio{s}$ &\\
     \hline
     & $2$ & $1$ & $5.5$ &\\
     & $3$ & $0.33$ & $5.15$ &\\
     & $4$ & $0.13 $ & $2.63$ & 
   \end{tabular}
  \end{ruledtabular}
  \caption{\label{tab:4}Values of $u$ and $<s>$ for values of $k_{max} = 2$, $3$ and $4$. We can see $u$ goes to zero, and $\promedio{s}$ tend to one if $k_{max}$ increase, 
which means the components are disappearing when $k_{max}$ is increased as observed in Figure \ref{fig:12}.}
 \end{center}
\end{table}

\noindent In table \ref{tab:4} we show some values of $u$ and $\promedio{s}$ calculated for $k_{max}$ values of $2$, $3$ and $4$. Where we can see how the mean component size and the $u$ value decrease, which means that components in the network are vanishing.

\section{Summary and Conclusions}

\noindent In this work we studied an asset exchange model with additive
or multiplicative exchanges. We consider the cases when bankruptcy is
introduced and also the interaction of agents through a social network. In the undirected fully connected network,   
is already known the additive case leads to a
\emph{Boltzmann - Gibbs} distribution, while in the multiplicative case, the system evolves to a condensed state where a single agent
gets all the money. When we allow for bankruptcy, condensation is
obtained for both types of exchange. We studied these condensation times, finding that the in the multiplicative
exchange the system condenses faster than in the  additive case.
The time of 
condensation is proportional to the number of agents. In the additive case 
the proportionality goes as $t_c^{add} \propto N^{1.899}$. For the multiplicative case the condensation time goes as 
$t_c^{mul}\propto N$.

\noindent When we introduce the social network we 
obtain stationary distributions in the additive case with a clear formation
of coexisting classes, the poor sector of the system following a
Boltzmann-Gibbs distribution until a value of wealth $2\promedio{w}$ and
a Pareto distribution from there on.

\noindent In the multiplicative case, there is also formation of four well
defined classes that arise from the presence of components in the network.
These components begin to disappear when the parameter $\promedio{k}$
is increased, since the connectivity of the network is higher. In this later
case we expect a global condensation to occur at a very long times.

\noindent Component formation and the absence of a giant component allows the presence of social
classes, since they prevent the money to flow in the entire network.
This has as a consequence that in some parts of the system the wealth can not be
larger than a maximum amount fixed by the number of members in the component,
effectively preventing global condensation.
The number of components increases as $\promedio{k}$ diminishes. In a real society, the flow of money is 
affected by a myriad of processes that cannot be accounted for in simple model like this. However, it is interesting to notice that despite its simplicity, the model can give some insights
about the origin of economic classes. As a conclusion, the isolation of agents
in a social network can lead to economic stratification, as observed in the
analysis of the present dynamics. For future work, it would be of interest to
analyze the same phenomena using different kinds of networks. 

\begin{acknowledgments}
\noindent This work has been partially supported by CONACyT-M\'exico and
PROMEP-M\'exico under grant NPTC-256.
\end{acknowledgments}

\newpage
\bibliography{paper}

\end{document}